\documentclass{article}
\usepackage{spconf}
\usepackage{amsmath,graphicx}
\usepackage{amssymb}
\usepackage[utf8]{inputenc}
\usepackage[T1]{fontenc}
\usepackage{multicol}
\usepackage{subcaption}

\title{Reliability-Aware Weighted Multi-Scale Spatio-Temporal Maps for  Heart Rate Monitoring}
\name{
Arpan Bairagi$^{\star}$ \qquad
Rakesh Dey$^{\star}$ \qquad
Siladittya Manna$^{\dagger}$ \qquad
Umapada Pal$^{\star}$
}

\address{
$^{\star}$ CVPR Unit, Indian Statistical Institute, Kolkata, India \\
$^{\dagger}$ Department of Computational and Data Sciences, Indian Institute of Science, Bengaluru, India
}

\begin{document}\ninept
%
\maketitle
\begin{abstract}
Remote photoplethysmography (rPPG) allows for the contactless estimation of physiological signals  from facial videos by analyzing subtle skin color changes. However, rPPG signals are extremely susceptible to illumination changes, motion, shadows, and specular reflections, resulting in low-quality signals in unconstrained environments. To overcome these issues, we present a \textbf{Reliability-Aware Weighted Multi-Scale Spatio-Temporal (WMST) map} that models pixel reliability through the suppression of environmental noises. These noises are modeled using different weighting strategies to focus on more physiologically valid areas. Leveraging the WMST map, we develop an SSL contrastive learning approach based on Swin-Unet, where positive pairs are generated from conventional rPPG signals and temporally expanded WMST maps. Moreover, we introduce a new \textbf{High-High-High (HHH) wavelet map} as a negative example that maintains motion and structural details while filtering out physiological information. Here, our aim is to estimate heart rate (HR), and the experiments on public rPPG benchmarks show that our approach enhances motion and illumination robustness with lower HR estimation error and higher Pearson correlation than existing Self-Supervised Learning (SSL) based rPPG methods.
\end{abstract}

\section{Introduction}
\label{sec:intro}

Remote PPG deals with non-contact estimation of different kinds of physiological signals, such as heart rate and SpO2 (peripheral capillary oxygen saturation), which can be analyzed from small colour variations on facial skin due to the blood volume pulse (BVP).
``rPPG" is contactless and captures small colour variation, but provides a poor Signal to Noise Ratio (SNR) when affected by different environmental noises.
Traditional signal-processing-based and region-based approaches~\cite{poh2010advancements, lewandowska2011measuring, de2013robust, de2014improved, wang2016algorithmic, pilz2018local} rely on handcrafted features, but do not perform well under unconstrained situations. Whereas, recent deep learning methods~\cite{ niu2020video, lu2021dual,dey2025sftt} provide robust solutions to obtain better rPPG signal by learning on direct videos or preprocessed spatio-temporal embeddings.

Still, these deep learning methods rely on large-scale annotated data, and acquisition of such large-scale data is challenging, which led us to the use of SSL methods~\cite{gideon2021way, yue2023facial,speth2023non, li2023contactless, savic2024rs, savic2025rs+}. SSL-based rPPG methods mainly concentrate on learning invariances based on contrastive learning through extensive data augmentations, while also using conventional pre-processing techniques such as Multi-Scale Spatio-Temporal (MST) map~\cite{niu2020video}, which do not model noises properly. Moreover, the current SSL-based rPPG approaches utilize generic negative augmentations that are either trivially separable or lack the capability to maintain realistic facial dynamics.
\par\noindent
To resolve the challenges mentioned above, our contributions are:
\begin{itemize}
    \item We have proposed an advanced pre-processing technique for weighting pixels based on their reliability for important physiological signals, the Weighted Multi-Scale Temporal (WMST) Map, which produces better SNR. To prove that our proposed map provides better SNR, we have conducted several experiments, and the results show that our proposed technique outperforms the existing MST map for different statistical testing.
    \item We have used a high-frequency wavelet-based negative, High-High-High (HHH) wavelet map for contrastive learning. Such HHH wavelet map models motion and noises, which are unimportant for rPPG. To our knowledge, such an HHH map has not been used earlier.
\end{itemize}

\section{RELATED WORKS}
\label{sec:rel_works}
Initial deep learning-based rPPG methods used to handle unconstrained situations far better than traditional methods, leading towards better performances. However, these deep learning-based models heavily relied on ground-truth physiological annotations to produce more feasible outcomes. To overcome this problem, recent works have investigated self-supervised learning (SSL) methods, which discourage the use of labeled data.

Gideon et al.~\cite{gideon2021way} were among the first few who proposed a contrastive learning methodology to predict meaningful physiological representations in the direct form of unlabeled facial videos. Following \cite{gideon2021way} Yue et al.~\cite{yue2023facial} extended this concept and created a self-supervised method that uses spatial and temporal consistency across various regions of the face to enhance robustness for physilogical enhancement. Speth et al.~\cite{speth2023non}, on the other hand, chose to completely steer clear of the contrastive learning framework and instead opted for a non-contrastive framework, where the model learns to enforce consistency between different views of the same video without the need to explicitly construct negative samples.

Li et al.~\cite{li2023contactless} proposed using traditional signals for helping the model to physiology-aware weights. Later, RS-rPPG~\cite{savic2024rs} and \cite{savic2025rs+} extended the idea proposed by~\cite{li2023contactless} by using different physiology-based traditional positives and negatives augmentations, leading to improved robustness in rPPG estimation.

In contrast, our work improves robustness by enhancing spatiotemporal signal representations through WMST map construction and wavelet-aware augmentation for a contrastive learning based self-supervised learning framework without relying on ground truth (GT) labels.

\section{PROPOSED METHODOLOGY}
\label{sec:prop_meth}
Our proposed framework consists of two key components: (1) addition of \emph{WMST map}, which models different unconstrained situations harmful for rPPG estimation, and (2) for contrastive learning, use of \emph{HHH wavelet map} that retains motion and structural patterns.
\subsection{Reliability-Aware Weighted MST Map Generation}
\label{subsec:preprocessing}
Remote PPG can be modelled by very small colour changes, making the signal highly sensitive to environmental noise. In order to handle these issues, we propose an improvement of the MST map~\cite{niu2020video}, the WMST map.

Following the idea of the MST map, each frame ($I$) of the video is converted into YUV and stacked with RGB, as YUV tends to work better for physiological signals~\cite{niu2019rhythmnet}. For each frame, we extract six regions of interest (ROIs): forehead, mouth, left cheek (two regions), and right cheek (two regions). A binary mask is created for each of the ROIs. From this set of ROIs, all possible non-empty subsets are created, and from each of these subsets, one value is pooled, resulting in better physiological signals~\cite{niu2020video}.

In rPPG, the reliability of a pixel depends on environmental noises such as reflection, sensor noise, shadow, and also the likelihood that the pixel is from the skin region. Unlike the MST map that uses mean pooling of all of the pixels within an ROI, we have modelled the reliability of a pixel using different well-established statistical and mathematical approaches for pooling.


We estimate the skin probability of each pixel, and we adopt a chrominance-based skin modelling~\cite{jones2002statistical}. From the $U$ and $V$ channel of the ROIs, a chrominance mean vector $\boldsymbol{\mu} = [\mu_U, \mu_V]^\top$ and a covariance matrix $\boldsymbol{\Sigma}$ is computed from $Z= [U, V]^\top$. 
Then the Mahalanobis distance is used to capture the skin weight 
\begin{equation}
w_{\mathrm{skin}} =
\exp\!\left(
-
(\mathbf{z}-\boldsymbol{\mu})^\top
\boldsymbol{\Sigma}^{-1}
(\mathbf{z}-\boldsymbol{\mu})
\right),\quad \mathbf{z}\in Z.
\end{equation}

Pixels with strong edge responses signify non-physiological textures or motion boundaries. We employ the Stationary Wavelet Transformation (SWT) to suppress noise and motion artifacts~\cite{kumar2021stationary}. Using SWT, we obtain the $(HL, LH, HH)$ subband of the luminance channel. The local edge energy at each pixel is then suppressed by
\begin{equation}
w_{\mathrm{edge}} = \exp(-\sqrt{HL^2 + LH^2 + HH^2}).
\end{equation}

\noindent
For identifying reflection and shadow HSV colour space is used, as bright and weakly saturated pixels are exhibited by specular highlights:
\begin{equation}
s_{\mathrm{spec}} = (1 - S) \cdot V,
\label{eq:spec}
\end{equation}
where, $S$ and $V$ are the channels of HSV.

Pixels affected by specular reflections are then softly suppressed using an exponential weighting function
\begin{equation}
w_{\mathrm{reflection}} =
\exp\!\left(
-\max\big(0, s_{\mathrm{spec}} - \tau_r \big)
\right),
\end{equation}
where $\tau_r$ is a reflection threshold and set to $0.3$.

Similarly, dark pixels caused by shadows or insufficient illumination are penalized based on the luminance channel $Y(x,y)$:
\begin{equation}
w_{\mathrm{shadow}} =
\exp\!\left(
-\max\big(0, (1 - Y) - \tau_s \big)
\right),
\end{equation}
where $\tau_s$ defines the shadow threshold and is set to $0.7$.

The values of $\tau_r$ and $\tau_s$ are determined based on the statistical properties of the HSV and luminance channels.

All of these weights are aggregated adaptively based on the variance of each weight (Equation~\ref{eq:weight_agg}). Variance can model the discriminative power of a reliability cue for separating reliable and unreliable pixels within ROIs. Whereas Entropy, on the other hand, is a measure of uncertainty and tends to be high even in homogeneously noisy areas, thus being less reliable as a measure of signal quality, and learned fusion would add training complexity and risk of overfitting. At the same time, variance provides a simple, robust, and deterministic reliability cue.

\begin{equation}
w(x,y)=\sum_{k}
\frac{\mathrm{Var}(w_k)}{\sum_j \mathrm{Var}(w_j)}\, w_k(x,y),
\label{eq:weight_agg}
\end{equation}
where, $k,j\in\{\text{chrom, noise, reflection, shadow}\}$.

Finally, we perform spatial ROI-wise ($r$) weighted mean pooling:
\begin{equation}
\hat{I}_{r} =
\sum_{(x,y)\in r} w(x,y)\, I_r(x,y).
\end{equation}

Though the WMST map considers a variety of reliability features, all the pre-processing steps are deterministic, non-iterative, and offline. Also, each weighting factor considers only linear pixel operations, and hence the total computational complexity is $\mathcal{O}(Height\times Width \times Time)$, which is the same as the standard MST map with a small constant factor. Thus, the proposed weighting approach enhances robustness without adding complexity to the model.

\begin{figure}[!t]
    \centering
    \includegraphics[width=0.8\linewidth]{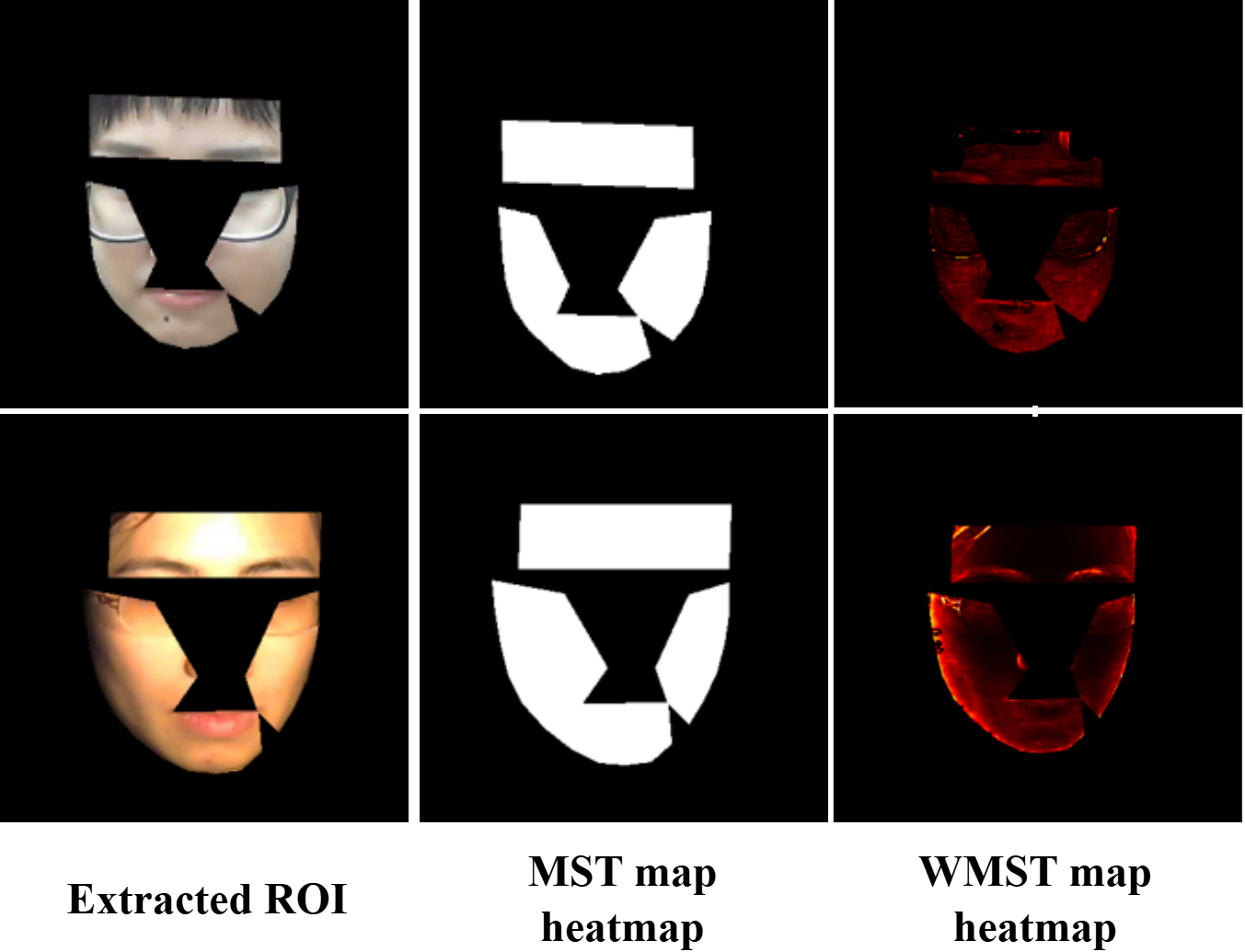}
    \caption{Qualitative comparison between the MST map and the proposed WMST map. Darker regions mean lower weights, and brighter regions mean higher weights (Best view by 300\% zoom).}
    \label{fig:heatmap_comp}
\end{figure}

A qualitative comparison is shown in Figure~\ref{fig:heatmap_comp}. Note that regions with hair, wrinkles, and regions with brighter reflection caused by unconstrained settings are handled and weighted properly by the proposed WMST map, whereas the MST map initializes weights to each region uniformly. 
\subsection{Contrastive Learning}
\label{sec:contrastive-learning}
We employ a contrastive learning framework to learn robust rPPG representations without using any physiological labels.

For contrastive learning, we have considered the WMST maps as our anchor. We try to pull the positive signals or the signals that are helpful for physiology close and push negatives or physiologically irrelevant signals far from the anchor.


Following previous works~\cite{savic2024rs, savic2025rs+}, we have constructed our positives and negatives. For positives, we use concatenation of 1D physiological signals obtained from video using six traditional approaches: ICA~\cite{poh2010advancements}, PCA~\cite{lewandowska2011measuring}, CHROM~\cite{de2013robust}, PVB~\cite{de2014improved}, POS~\cite{wang2016algorithmic}, and LGI~\cite{pilz2018local}. In addition to output signals from traditional models, temporally resized WMST maps with random ROI shuffling are used as augmentations along with traditional signals. As negatives, we consider WMST maps from different videos and temporally augmented WMST maps from different videos.

\subsubsection{High-High-High (HHH) Map}
In addition to these common negatives, we propose a new HHH map to model motion, facial structure, and noise explicitly. High-frequency information mainly represents abrupt spatial and temporal variations, which are mostly irrelevant to the rPPG signal. To obtain the HHH map, we utilize 3D SWT on the input video $V \in \mathbb{R}^{H \times W \times T \times C}$ and obtain the HHH subband $V_{hhh}$, representing high-frequency variations in height, width, and time. After the same ROI decomposition as the MST map extraction, we perform spatial pooling for each ROI using root-mean-square aggregation to highlight high-frequency regions with strong activations:
\begin{equation}
\hat{I}^{HHH}_{r}(t) = \sqrt{\frac{1}{|r|} \sum_{(x,y)\in r} V_{hhh}(x,y,t)^2}.
\end{equation}
The HHH map retains motion and structural noise but removes physiological information, making it a difficult and useful negative sample for contrastive learning.

\subsubsection{Contrastive Loss}
To find the physiological signal between similar signals, we have used temperature-free InfoNCE loss~\cite{kim2025temperature} as our objective function.

Let the predicted signals be $\mathbf{y}$, $\{\mathbf{y}^{+}_i\}$, and $\{\mathbf{y}^{-}_k\}$ for anchor WMST map, positive samples, and negative samples, respectively. These predicted signal transformed into the frequency domain, where a Power Spectral Density function (PSD) $\mathcal{P}(\cdot)$ is applied within the frequency band of $[0.5,3.0]$ Hz~\cite{savic2024rs} and then normalized.
For finding the similarity between two signals, we have utilized the Pearson correlation coefficient $r$ to focus on the  shape and phase of the signal:
\begin{equation}
r(\mathcal{P}(\mathcal{Y}_a), \mathcal{P}(\mathcal{Y}_b)) = \frac{
\left(\mathcal{P}(\mathcal{Y}_a) - \overline{\mathcal{P}(\mathcal{Y}_a)}\right) \cdot \left(\mathcal{P}(\mathcal{Y}_b) - \overline{\mathcal{P}(\mathcal{Y}_b)}\right)
}{
\left\|\mathcal{P}(\mathcal{Y}_a) - \overline{\mathcal{P}(\mathcal{Y}_a)}\right\|_2 \left\|\mathcal{P}(\mathcal{Y}_b) - \overline{\mathcal{P}(\mathcal{Y}_b)}\right\|_2
}.
\end{equation}
Since the classical InfoNCE loss suffers from the vanishing gradient and slower convergence, we have employed a temperature-free InfoNCE loss, which uses $2.0 \cdot arctanh$ on the similarity~\cite{kim2025temperature}, and gets rid of additional hyperparameter tuning.
\begin{equation}
h(\mathbf{y}_a, \mathbf{y}_b) = 2.0 \cdot \operatorname{arctanh}\left( r(\mathcal{P}(\mathbf{y}_a), \mathcal{P}(\mathbf{y}_b)) \right),
\end{equation}

The final contrastive learning objective is formulated as:
\begin{equation}
\mathcal{L}_{\mathrm{nce}} =
- \log \left(
\frac{
\exp(h(\mathbf{y}, \mathbf{y}^{+}))
}{
\sum\limits_{k}\exp(h(\mathbf{y}, \mathbf{y}^{-}_k))
}
\right),
\end{equation}
\subsection{Network Architecture and Optimization}
\label{subsec:network}
\subsubsection{Backbone Architecture}
We have used Swin-Unet~\cite{cao2022swin} as our backbone, as it uses self-attention to model long-range temporal dependencies, being also successful for the rPPG signal~\cite{savic2024physu}.

We have modelled our task as a spatial-temporal map to a spatial-temporal map construction problem instead of directly regressing the HR, i.e., WMST maps are fed into the Swin-Unet and in output a refined spatio-temporal signal is constructed of the same shape as our input using contrastive learning.


\subsubsection{Pretraining Strategy}
We first train our network on signals generated by traditional approaches described in Section~\ref{sec:contrastive-learning} to learn physiologically relevant weights for the model. For this stage, we have used a linear combination of normalized $L_1$ distance and Pearson correlation coefficient of predicted and traditional signals computed within the physiological band of $[0.5,3.0]$ Hz \cite{savic2024rs} to minimize the distance and ensure temporal alignment properly. These 1D signals, which are generated from traditional approaches, are similar features maps like the WMST or MST maps.
The total pretraining loss is formulated as:
\begin{equation}
\mathcal{L}_{\text{pre}} = \left\| {\mathcal{P}}(y) - {\mathcal{P}}(y_{t}) \right\|_1 + \lambda \cdot \left( 1 - r({\mathcal{P}}(y), {\mathcal{P}}(y_{t}))\right),
\end{equation}
where $\lambda$ is a scaling hyperparameter set to 0.05. A small $\lambda=0.05$ enables the model to focus on the correct frequency distribution first and then refine the temporal alignment.
\subsubsection{Heart Rate Estimation}
\begin{equation}
\mathrm{HR} = 60 \times \arg\max_{f \in [0.5, 3.0]} P(f).
\label{eq:hr}
\end{equation}
The predicted spatio-temporal signal map is averaged over channels and ROIs to obtain a one-dimensional temporal signal, which is band-pass filtered in the physiological range $[0.5, 3.0]$~Hz~\cite{savic2024rs}. HR is then estimated from the dominant frequency of its PSD (Equation~\ref{eq:hr}).




\section{Experiments and Results}
\label{sec:experiments_and_results}
In order to evaluate the significance of our proposed idea of the WMST map, we have shown the statistical significance analysis and results on different datasets.
\subsection{Dataset}
Our approach is tested on two public benchmark datasets, VIPL-HR~\cite{niu2018vipl} and UBFC-rPPG~\cite{bobbia2019unsupervised}. Following previous works, we use same protocols for obtaining the results: 5-fold for VIPL-HR, and 30-second segments for UBFC-rPPG.

VIPL-HR is a challenging dataset with 2,378 RGB videos of 20-30 seconds long from 108 subjects, shot under different illumination conditions, motion types, and settings. Whereas, UBFC-rPPG is a dataset of one-minute RGB videos of 42 subjects under stable illumination with less motion.

These two datasets allow us to test the proposed approach in both unconstrained and controlled settings.
\subsection{Statistical Significance Analysis}
\label{subsec:stat_analysis}
We use a one-sided test on the VIPL-HR dataset to establish that our proposed method significantly enhances SNR. In a one-sided test, our null hypothesis ($H_0$) states that there is no improvement in SNR (Equation~\ref{eq:null}), whereas the alternative hypothesis ($H_1$) states that the WMST map provides higher SNR (Equation~\ref{eq:alt}).

\begin{equation}
        \Delta_i = \mathrm{SNR}^{\mathrm{WMST}}_i - \mathrm{SNR}^{\mathrm{MST}}_i,
        \label{eq:snr}
\end{equation}
%
\begin{equation}
H_0:\ \mathbb{E}[\Delta_i] \le 0,
\label{eq:null}
\end{equation}%
\begin{equation}
H_1:\ \mathbb{E}[\Delta_i] > 0.
\label{eq:alt}
\end{equation}

For one-sided paired tests, we have considered the t-test, which works under the normality assumption. Since the number of paired samples is greater than 50, the test is valid even if the data are not normally distributed, as explained by the Central Limit Theorem~\cite{lumley2002importance}.

\noindent\textbf{Investigating Robustness to Outliers:} Along with the t-test, we have also conducted the Wilcoxon signed-rank test to prove the robustness of our method to outliers.

The Wilcoxon signed-Rank test strongly rejects $H_0$ with extremely small $p$-values, confirming that the proposed WMST map significantly improves SNR over the MST map, even under possible skewness of the underlying distribution, and presence of outliers (Table \ref{tab:snr_stats}).

\begin{figure}[!t]
    \centering
    \includegraphics[width=1.0\linewidth]{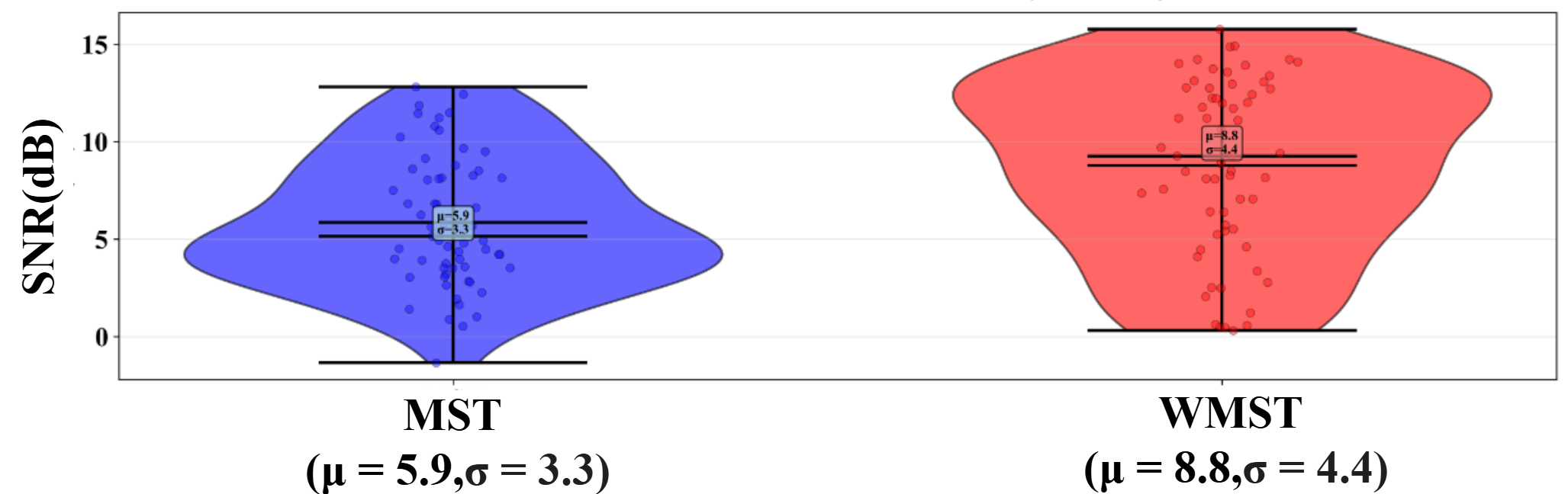}
    \caption{Distribution of SNR of G channel values across all ROI combinations obtained from all non-empty subsets from the set of six ROIs.}
    \label{fig:violin_snr}
\end{figure}

In addition to the statistical tests, the violin plots in Figure~\ref{fig:violin_snr} clearly indicate that there is a positive shift in the distribution of the SNR values when employing the proposed WMST map over the traditional MST mapping. The increased mean value and spread of the distribution indicate better separation of reliable and unreliable ROIs.

\begin{figure}[t]
    \centering
    \begin{subfigure}[t]{\linewidth}
        \centering
        \includegraphics[width=\linewidth]{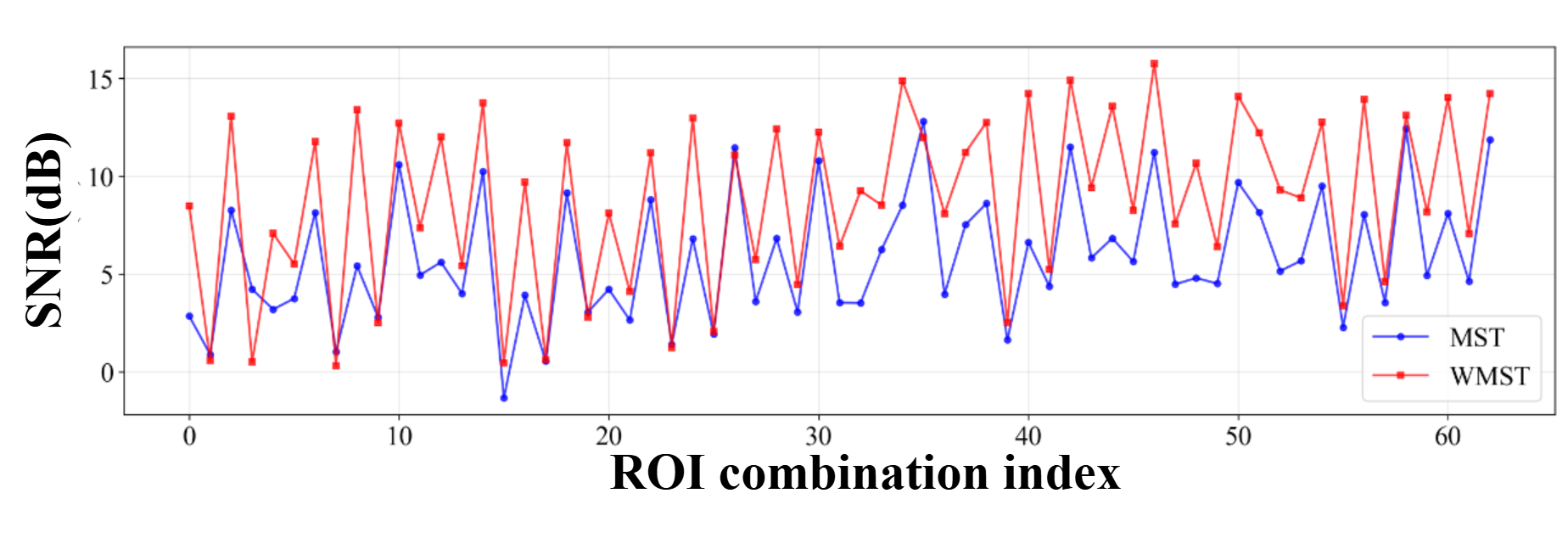}
        \caption{SNR values across ROI pairs averaged over frames and channels.}
        \label{fig:signal_comp_snr}
    \end{subfigure}
    
    \begin{subfigure}[t]{\linewidth}
        \centering
        \includegraphics[width=\linewidth]{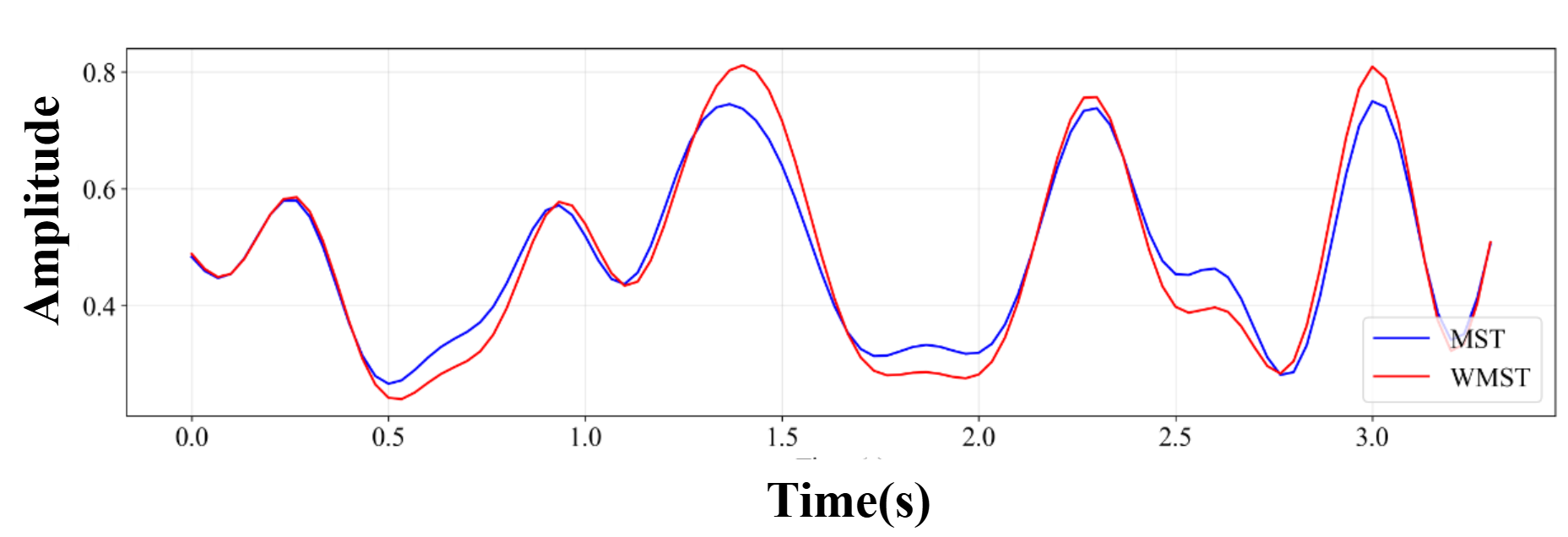}
        \caption{Temporal signal averaged over all ROIs and channels.}
        \label{fig:signal_comp_temporal}
    \end{subfigure}

    \caption{Qualitative comparison of signal quality between the MST map and the proposed WMST map (Best viewed with 300\% zoom).}
    \label{fig:signal_comp}
\end{figure}

Further, Figure~\ref{fig:signal_comp} illustrates that the proposed reliability-weighting approach results in not only higher SNR values across all combinations of ROIs but also a smoother averaged temporal signal.
\begin{table}[ht]
\centering
\caption{One-sided statistical significance analysis of SNR on VIPL-HR (WMST $>$ MST).}
\label{tab:snr_stats}
\begin{tabular}{ccc}
\hline
Setting & Paired $t$-test & Wilcoxon signed-rank\\
\hline
WMST vs MST&$1.90\times10^{-13}$&$3.18\times10^{-13}$\\
\hline
\end{tabular}
\end{table}

\subsection{Results}
As mentioned earlier, the proposed framework is tested on two public benchmark datasets, VIPL-HR~\cite{niu2018vipl} and UBFC-rPPG~\cite{bobbia2019unsupervised}. 
\subsubsection{Quantitative Results}
Detailed quantitative results which are produced by our proposed method are reported and compared in terms of mean absolute error (MAE), root mean squared error (RMSE), and Pearson correlation coefficient (R).


The results on the VIPL-HR dataset are shown in Table~\ref{tab:vipl_hr}. The proposed method shows the \textbf{best MAE} and \textbf{highest Pearson correlation} among all state-of-the-art (SOTA) SSL methods. Although RS+rPPG~\cite{savic2025rs+} shows a slightly better RMSE, our method clearly shows better accuracy in terms of MAE and better synchronization with the GT, as shown by the highest Pearson correlation coefficient. These experiments demonstrate that the proposed method not only provides a better estimation of heart rate but also models the signal dynamics better for one of the most difficult datasets.

The experimental results on the UBFC-rPPG dataset are provided in Table~\ref{tab:ubfc}. The proposed method is comparable to the SOTA SSL methods and has a similar MAE and RMSE value while maintaining a high Pearson correlation, like other SOTA methods. This is an indication that the learned features are generalizable to a controlled setting, although the method is primarily designed for handling noisy and motion artifacts.



The results on both datasets demonstrate that the proposed WMST map and wavelet-aware contrastive learning framework enhance the heart-rate estimation robustness.

\begin{table}[ht!]
\centering
\caption{Performance comparison on the VIPL-HR dataset with SOTA SSL methods. The best and second best result is denoted by bold and underline, respectively.}
\label{tab:vipl_hr}
\begin{tabular}{cccc}
\hline
Method & MAE$\downarrow$ & RMSE$\downarrow$ & $R\uparrow$ \\
\hline
Contrast-Phys~\cite{sun2024contrast} & 17.8 & 22.8 & 0.17 \\ \hline

SSPD~\cite{zhang2024self}& 6.04 & 9.56 & 0.59 \\ \hline
Li et al.~\cite{li2023contactless}& 5.19 & 8.26 & 0.78 \\ \hline
Greip et al.~\cite{zhang2025advancing}& 7.35 & 9.70 & 0.55 \\ \hline
RS-rPPG.~\cite{savic2024rs}& 5.98 & 10.5 & 0.56 \\ \hline
RS+rPPG~\cite{savic2025rs+}& \underline{4.40} & \textbf{7.19} & \underline{0.79} \\ \hline
\textbf{Ours}& \textbf{4.32} & \underline{7.26} & \textbf{0.86} \\
\hline
\end{tabular}
\end{table}

\begin{table}[ht!]
\centering
\caption{Performance comparison on the UBFC-rPPG dataset with SOTA SSL methods. The best and second best result is denoted by bold and underline, respectively.}
\label{tab:ubfc}
\begin{tabular}{cccc}
\hline
Method & MAE$\downarrow$ & RMSE$\downarrow$ & $R\uparrow$ \\
\hline
Gideon et al.\cite{gideon2021way}& 2.3 & 2.9 & \textbf{0.99} \\ \hline
SiNC~\cite{speth2023non} & 0.59 & 0.83 & \textbf{0.99} \\ \hline
Contrast-phys~\cite{sun2024contrast} & 1.00 & 1.40 & \textbf{0.99} \\ \hline
Yue et.al.~\cite{yue2023facial} & 0.58 & \underline{0.84} & \textbf{0.99} \\ \hline
Li et al.~\cite{li2023contactless}& \textbf{0.48} & \textbf{0.64} & \textbf{0.99} \\ \hline
RS+rPPG~\cite{savic2025rs+}& 0.62 & 0.99 & \textbf{0.99} \\ \hline
\textbf{Ours}& \underline{0.50} & 0.86 & \textbf{0.99} \\
\hline
\end{tabular}
\end{table}
\subsubsection{Ablation Study}
We also perform an ablation study on the VIPL-HR dataset to analyze the role of the proposed WMST map and the HHH wavelet map augmentation strategy. The results of the ablation study are presented in Table~\ref{tab:ablation} in terms of MAE and Pearson correlation. With the WMST map incorporated into RS-rPPG, there is a visible error correction, which proves that the reliability-aware spatial weighting is beneficial for improving the quality of the extracted signals. Without the HHH map in our full model, the Pearson correlation value decreases, which means that the absence of the explicit negative samples for motion and structure has an influence on the synchronization of the signals with the GT.
\begin{table}[!ht]
\centering
\caption{Ablation study on the VIPL-HR dataset.}
\label{tab:ablation}
\begin{tabular}{ccc}
\hline
\textbf{Setting} & \textbf{MAE} $\downarrow$ & $R$ $\uparrow$ \\
\hline
RS-rPPG~\cite{savic2024rs} + WMST map     & 6.01 & 0.65 \\
\hline
\textbf{Ours} w/o HHH map     & 4.78 & 0.80 \\
\hline
\textbf{Ours} w/o WMST map    & 4.55 & 0.81 \\
\hline
\textbf{Ours with HHH map and WMST map}        & \textbf{4.32} & \textbf{0.86} \\
\hline
\end{tabular}
\end{table}
\subsubsection{Qualitative Results}
\begin{figure}[ht!]
    \centering
    \begin{subfigure}[!t]{0.9\linewidth}
        \centering
        \includegraphics[width=\linewidth]{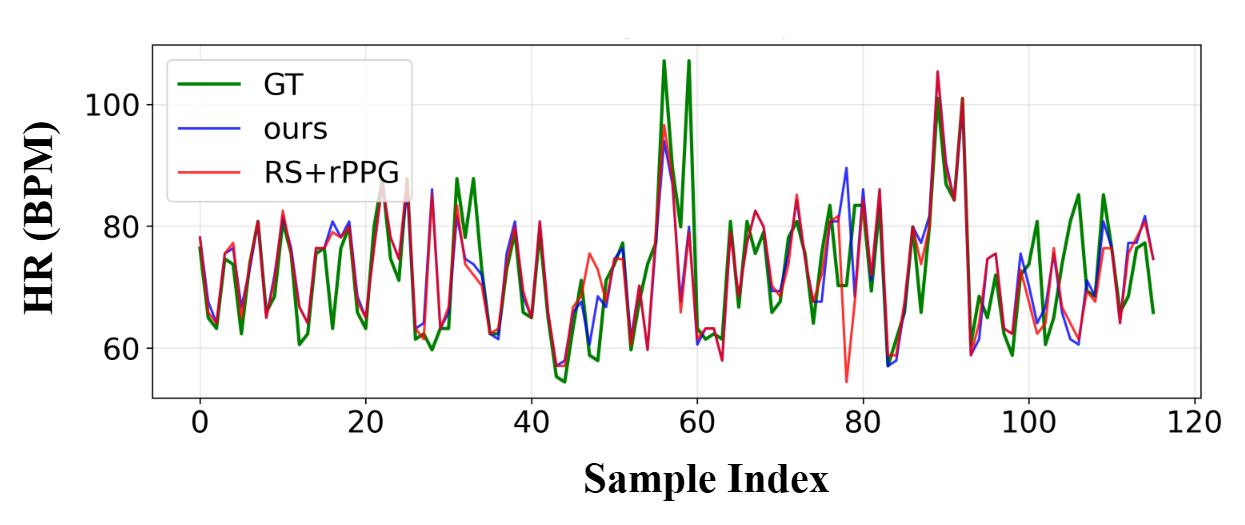}
        \caption{Prediction of GT HR vs Proposed vs RS+rPPG for Segment 1.}
        \label{fig:hr_comp_1}
    \end{subfigure}
    
    
    \begin{subfigure}[t]{0.9\linewidth}
        \centering
        \includegraphics[width=\linewidth]{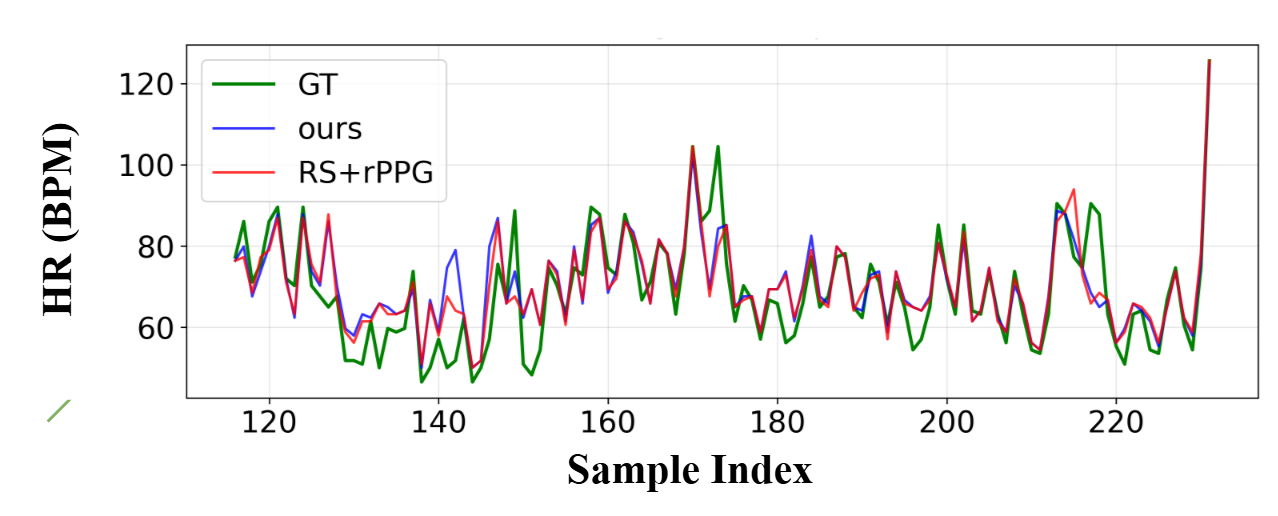}
        \caption{Prediction of GT HR vs Proposed vs RS+rPPG for Segment 2.}
        \label{fig:hr_comp_2}
    \end{subfigure}
    
    
    \begin{subfigure}[t]{0.9\linewidth}
        \centering
        \includegraphics[width=\linewidth]{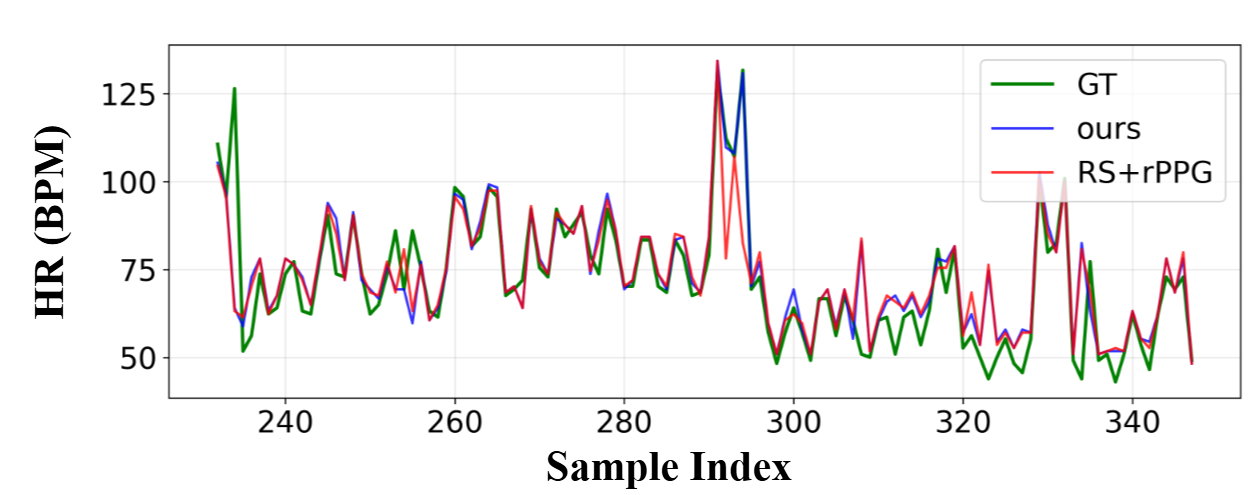}
        \caption{Prediction of GT HR vs Proposed vs RS+rPPG for Segment 3.}
        \label{fig:hr_comp_3}
    \end{subfigure}
    
    
    \begin{subfigure}[t]{0.9\linewidth}
        \centering
        \includegraphics[width=\linewidth]{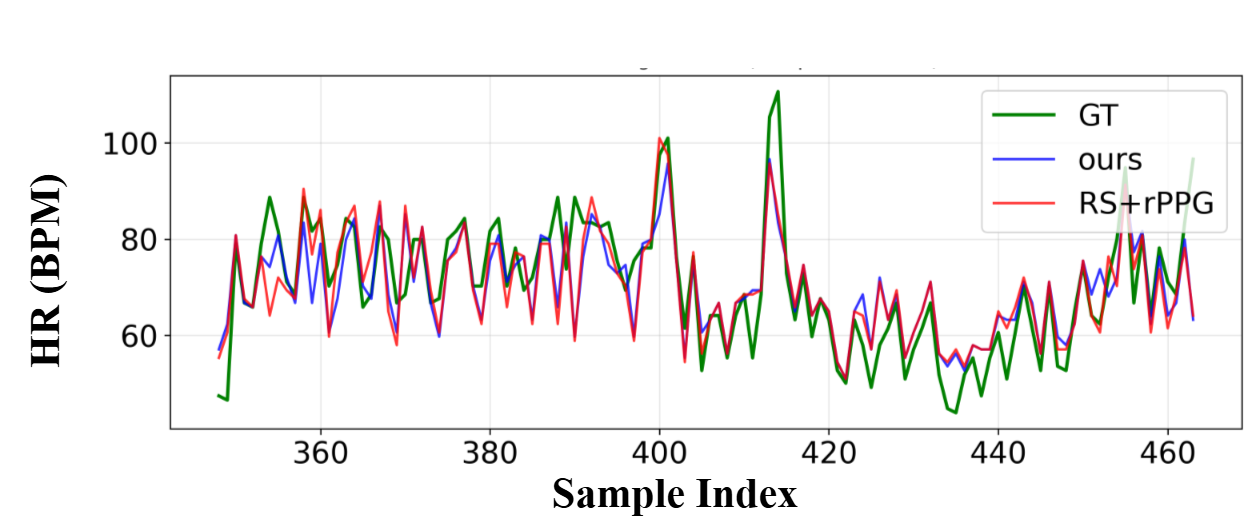}
        \caption{Prediction of GT HR vs Proposed vs RS+rPPG for Segment 4.}
        \label{fig:hr_comp_4}
    \end{subfigure}

    \caption{Qualitative comparison of predicted HR predictions between RS+rPPG and the proposed method on the VIPL-HR dataset. We have segmented all of the predictions into four segments for ease of visualization (Best viewed with 300\% zoom).}
    \label{fig:heartrate-comparison}
\end{figure}
Figure~\ref{fig:heartrate-comparison} shows a qualitative comparison of heart rate predictions on the VIPL-HR dataset, where the entire sequence is split into four successive segments. The proposed approach outperforms the SOTA in several critical areas: in Segment 1 (Figure~\ref{fig:hr_comp_1}), our approach more accurately models the double-peak pattern around samples 50-65, while RS+rPPG displays spurious oscillations around samples 90-105. In Segment 2 (Figure~\ref{fig:hr_comp_2}), RS+rPPG displays severe overshoots, while our approach keeps a much closer track of the GT. In Segment 3 (Figure~\ref{fig:hr_comp_3}), both approaches correctly track the sharp peak around sample 295, while our approach keeps a much steadier track of the samples that follow (300-347). In Segment 4 (Figure~\ref{fig:hr_comp_4}), both approaches perform equally well. In general, our approach displays fewer spurious peaks and a more consistent tracking pattern, which justifies the higher Pearson correlation scores reported in our quantitative analysis.

\subsection{Training Setup}
For both the pre-training and contrastive learning stages, we have utilized Swin-Unet. In the pre-training stage, we have used the AdamW optimizer with a learning rate of $1e-5$, epsilon of $1e-8$, betas $(0.9, 0.99)$, weight decay of $0.001$, and batch size of $8$. In order to reduce overfitting, we have reduced the learning rate when the loss plateaus or does not decrease for $10$ epochs with a factor of $0.1$. In both the pre-training and contrastive learning stages, the model trains for $30$ epochs.

\section{Conclusion}
\label{sec:conclusion}
This paper presents an SSL framework with Reliability-Awareness for remote physiological monitoring under unconstrained settings. We introduce the WMST map, which explicitly models environmental noises to suppress physiologically weak regions. In addition, the HHH wavelet map was proposed as an informative negative sample, enabling improved separation between physiological signals and motion-induced noise in contrastive learning. Experimental results on public benchmarks demonstrate that the proposed approach achieves improved accuracy and stronger physiological consistency compared to existing SSL methods. Future work will explore alternative reliability fusion strategies for the WMST map and evaluate cross-dataset generalization and mobile deployment.

{
\bibliographystyle{IEEEbib}
\bibliography{refs}
}
\end{document}